# College Dropout Factors: An Analysis with LightGBM and Shapley's Cooperative Game Theory

Hugo Roger Paz

Professor and researcher of the Faculty of Exact Sciences and Technology of the National University of Tucumán.
PhD candidate in Exact Sciences and Engineering at the Faculty of Exact Sciences and Technology of the National University of Tucumán.
Email: hpaz@herrera.unt.edu.ar
ORCID: https://orcid.org/0000-0003-1237-7983

**Abstract**
This study was based on data analysis of academic histories of civil engineering students at FACET-UNT. Our main objective was to determine the academic performance variables that have a significant impact on the dropout of the career. To do this, we implemented a correlation model using LightGBM (Barbier et al., 2016; Ke et al., 2017; Shi et al., 2022). We use this model to identify the key variables that influence the probability of student dropout.

In addition, we use game theory to interpret the results obtained. Specifically, we use the SHAP library (Lundberg et al., 2018, 2020; Lundberg & Lee, 2017) in Python to calculate the Shapley numbers.

The results of our study revealed the most important variables that influence the dropout from the civil engineering career. Significant differences were identified in terms of age, time spent in studies, and academic performance, which includes the number of courses passed and the number of exams taken. These results may be useful to develop more effective student retention strategies and improve academic success in this discipline.

## 1.- Introduction

Academic dropout in higher education institutions represents a crucial challenge in the quest for quality education and equity. This phenomenon not only has a significant impact on the lives of students, but also raises important concerns for educational institutions and society in general (Cabrera et al., 2006; Tinto, 1975). Identifying the variables that influence academic dropout and its relationship with academic performance is essential for effective student retention measures. In this context, this study focuses on analysing the academic histories of civil engineering students at FACET-UNT, with the aim of determining the academic performance variables that have a significant impact on student dropout. Addressing this problem is essential to improve the effectiveness of retention strategies.

To gain an understanding of this problem and its relationship to academic performance, the following objectives are set out:

- Identify variables with the greatest impact: The first objective is to identify the academic performance variables that have a significant impact on dropout. This involves determining which specific aspects of the students' academic trajectory, such as the number of courses passed, the time spent in studies, among others, are most significantly associated with the probability of dropout.

- Applying Game Theory: The second objective is to incorporate game theory into the analysis of academic attrition. This involves using this analytical perspective to understand how individual students' decisions may be influenced by the actions of others, and how this may affect dropout rates.

- Interpreting the Results with SHAP: The third objective focuses on interpreting the results of the LightGBM model using the SHAP library in Python. This step will allow us to understand the relative contribution of each variable to the students' decision-making process and thus to academic attrition.

Taken together, these objectives are designed to provide an understanding of the dynamics of academic attrition in the context of civil engineering students. The results of this research have the potential to inform more effective student retention strategies.

2.- Method

**Population and sample**

The data under analysis come from the academic trajectories' of 1615 students enrolled in the Civil Engineering Major, covering a diverse spectrum of progress in the major, from first-year students to those in stages close to graduation, as well as individuals who have discontinued their education and those who have completed their educational process. These data were obtained through the Student and Administrative Management System (SIGEA in Spanish), which collects and centralizes relevant information on the academic progress and management of students in the major.

In the framework of this research, we examined the academic performance of university students over a period of fourteen years, from 2004 to 2019. To ensure the accuracy of the results, an exclusion was made for those students who had obtained the approval of subjects through a system of equivalencies, due to modifications in their curriculum, academic path or educational institution. This measure was implemented considering that these students may have entered the major with subjects already passed, which could impact their progress and performance in comparison with their peers. For the analysis, a set of 1343 academic records were selected from a total of 1615.

In the present context, students who have completed their studies within the period under evaluation (2005-2019) are classified as "graduates". On the other hand, students who have remained enrolled in the academic programme until 2019 and have not yet completed their studies are considered as "permanent".

Ultimately, "drop-out" status is defined for those students who do not register any enrolment in 2020 and have not completed their educational process. In adopting this approach, we follow the proposal of

González Fiegehen (2007) for the students who enrolled in 2019, representing only 5% of the total analysed (Paz, 2022). For the rest of the students, a less rigorous criterion is used, including those who re-enrolled in the course and resume their studies after a period of inactivity. The calculation of the time lapses was based on the difference between the last activity documented in the SIGEA system and the date of entry to the academic programme.

The five-and-a-half-year study programme, spread over a total of eleven semesters, follows a Correlative System that establishes necessary conditions for enrolment in courses. These conditions can be "regularised", which means that partial evaluations or practical tasks have been passed, or "passed", which indicates that the final exam of the course has been successfully completed, thus guaranteeing its approval.

For the analysis, academic records were used (see section 2.1.), from which the following data were extracted for 1343 students:

- Genre
- Student's age at last recorded activity (EdadUltimaActividad)
- Time spent at the faculty (TiempoFacultad)
- Student Dropout (Abandono)
- Average of Examination Grades (PromedioNotas)
- Total number of courses taken-includes those with resources (NumeroTotalCursadas)
- Number of courses in which you passed the Final Examination (NumeroRegulares)
- Number of courses retaken (NumeroRecursadas)
- Number of courses in which he/she could not access the Final Examination (NumeroLibres)
- Total number of Final Examinations (NumerodeExamenes)
- Number of courses passed during the course (NumeroPromociones)
- Number of Approved Courses (NumeroAprobados)
- Number of failed courses (NumerodeReprobados)
- Number of Final Examinations not attended (NumerodeAusentes)

For greater understanding, its name in Spanish is included in parentheses as it appears in the corresponding graphics.

**2.2.- Analysis technique: LightGBM**

Choosing the right tools and methodologies in research is a crucial process that directly impacts the quality and effectiveness of the results obtained. In the case of this study, the selection of LightGBM as the machine learning algorithm and the integration of game theory is based on several fundamental reasons:

- Computational Efficiency: LightGBM is widely recognised in the machine learning community for its computational efficiency and its ability to handle large and complex datasets. Its implementation of leaf-wise tree growth allows for a faster training process compared to other tree-based algorithms, which is especially beneficial when analysing a large volume of student academic history data.

- High Performance: LightGBM is designed to provide high performance in terms of prediction accuracy. This is essential when addressing classification and prediction problems in the context of academic attrition, where the aim is to identify patterns and determinants as accurately as possible.

- Interpretability: Despite its complexity, LightGBM offers tools and methods to interpret and understand the model results. This is crucial for our goal of identifying the academic performance variables most influential in student dropout.

- Game theory: Incorporating game theory into the analysis adds an additional analytical approach to understanding the dynamics behind students' decisions to drop out. This theory allows us to consider how a student's actions can be influenced by the actions of others, which can have significant implications for academic decision-making.

- Holistic Approach: The combination of LightGBM and game theory provides us with a holistic approach to address the problem of academic dropout. By using an advanced machine learning model together with game theory, we are better positioned to explore and understand the complex interactions between academic performance variables and student decision factors.

Overall, the choice of LightGBM and game theory is justified by their ability to effectively address the problem of academic attrition, offering a combination of accuracy, efficiency and interpretability that supports our research objectives.

The process is detailed below, including the parameters used in the application of the model:

**Division of the Data:**
The modelling process was started by splitting the data into training and test sets. For this purpose, the `train_test_split` function of the scikit-learn library was used. (Pedregosa et al., 2011) with a proportion of 70% of the data for training and 30% for testing. This split allows to evaluate the performance of the model on data not used during training and to ensure its generalisability.

**Creation of LightGBM Datasets:**
Next, LightGBM-specific datasets, named `d_train` and `d_test`, were created using the `lgb.Dataset` class. These sets were constructed from the previously split training and test data.

**Definition of Model Parameters:**
The parameters of the LightGBM model were carefully defined to achieve a balance between performance and regularisation. The parameters used are as follows:

- `max_bin`: 512
- `learning_rate`: 0.05

- boosting_type: "gbdt": `boosting_type: "gbdt"`.
- objective: "binary": `objective: "binary"`.
- metric: "binary_logloss": `"binary_logloss"`.
- num_leaves: 10
- verbose: -1
- min_data: 100
- boost_from_average: True

These parameters are set according to LightGBM recommendations and based on previous experience in hyperparameter selection. The aim is to optimise model accuracy and control complexity.

**Training of the Model:**
Finally, the LightGBM model was trained using the `lgb.train` function. The training data sets were specified and the number of iterations (in this case, 10,000) was given. In addition, the performance on the test set was evaluated using the `valid_sets=[d_test]` option.

It is important to note that, although not included in this example, in a production environment it is common to use techniques such as cross-validation and hyperparameter search to further refine the model and ensure its robustness.

This approach used in the application of LightGBM allows reliable results to be obtained that are comparable with research standards in internationally renowned journals.

**2.3.- Analysis technique: Shapley's Cooperative Game Theory.**

In this study, we have incorporated game theory as a valuable analytical framework for understanding the dynamics behind students' decisions regarding academic dropout. Before delving into how game theory was applied in our analysis, it is important to provide a brief explanation of the key concepts and tools used in this methodology.

Game theory is a field of study that focuses on the analysis of strategic decisions made by rational agents when their actions affect both their outcomes and the outcomes of other agents. In the context of higher education, game theory can be used to model how individual students' decisions, such as whether to drop out or continue their studies, can be influenced by the decisions of other students and contextual factors.

Shapley numbers are a fundamental tool in cooperative game theory that is used to assign a numerical value to the relative contribution of each player in a cooperative game. In the context of our analysis, Shapley numbers are used to understand how much each variable (e.g. academic performance) contributes to students' decision making with respect to academic dropout.

To incorporate game theory into our analysis, we applied the SHAP (SHapley Additive exPlanations) library, which allows us to calculate Shapley numbers and visualise the contribution of variables to the LightGBM model. How this process was carried out is detailed below:

**Calculation of Shapley Numbers:**
We use the `shap.TreeExplainer` function to create an explainer specific to the LightGBM model. This explainer allows us to calculate the Shapley values for each variable in our model.

**Visualisation of the results:**
The following Shap tools were used to visualise the results:

- Summary Plot: We create a summary plot using the `shap.summary_plot` function. This plot provides an overview of the importance of each variable in the model and how it affects the overall predictions.

- Dependency Plot: We then generate dependency plots for each individual variable using a loop. These plots show how variation in each variable influences the model's predictions.

- SHAP Interaction Values: We calculate SHAP interaction values using the `shap.TreeExplainer` function to assess how interactions between variables may influence model predictions.

- SHAP Interaction Value Dependence Plots: Finally, we create SHAP interaction value dependence plots to explore how interactions between two variables affect model predictions. These plots allow for a deeper analysis of the relationships between variables.

## 3. Results

In this study, we have applied several evaluation metrics to measure the performance of our LightGBM model and to assess the effectiveness of our predictions in the context of academic attrition in civil engineering students at FACET-UNT. These metrics are standard in the evaluation of binary classification models and provide a comprehensive understanding of how our model performs. In the following, we discuss the metrics used:

Accuracy: the proportion of correct predictions.
Precision: the proportion of true positives among all positive predictions.
Recall: the proportion of true positives among all true positive cases.
F1 Score: a metric that combines precision and recall.
ROC AUC Score: the area under the ROC (Receiver Operating Characteristic) curve.
Log Loss: the binary logarithmic loss.

The results are as follows.
Accuracy: 0.7816
Accuracy: 0.7957
Recall: 0.8170
F1 Score: 0.8062
ROC AUC Score: 0.8699
Log Loss: 0.6720

The results of the evaluation metrics for the LightGBM model indicate a solid performance in the task of academic dropout classification in civil engineering students at FACET-UNT. With an accuracy value of 0.7957, the model shows a high ability to correctly identify students who will drop out, thus minimising false positives. In addition, the recall value of 0.8170 indicates that the model is effective in capturing a large proportion of real dropout cases, which is critical for the early detection of at-risk students. The F1 Score of 0.8062, which combines accuracy and recall, suggests an appropriate balance between accurate identification and error minimisation. The high value of the area under the ROC curve (ROC AUC Score) of 0.8699 indicates that the model has a good ability to discriminate between positive and negative classes. Finally, the binary log loss value of 0.6720 is consistent with a well-calibrated model. Taken together, these results suggest that the LightGBM model is accurate in identifying academic attrition and shows robust performance in this specific context.

Table 1. Shapley values for importance variables

| Variable Shapley Number |
|---|
| NumeroPromociones_shap 3.651 |
| NumeroRegulares_shap 2.867 |
| PromedioNotas_shap 2.569 |
| TiempoFacultad_shap 2.533 |
| EdadUltimaActividad_shap 2.510 |
| NumeroRecursadas_shap 1.923 |
| NumeroTotalCursadas_shap 1.873 |
| MaxRegAcum_shap 1.726 |
| NumeroLibres_shap 1.658 |
| NumeroAprobados_shap 1.636 |
| NumerodeReprobados_shap 1.490 |
| NumerodeAusentes_shap 1.117 |
| NumerodeExamenes_shap 0.786 |
| Genero_shap 0.578 |

The results obtained by applying Shapley's game theory to analyse the incidence of variables on student dropout are illuminating in terms of the relative contribution of each variable to students' decision making on whether to drop out or continue their studies. Key observations based on the Shapley numbers are presented here:

- Shap_Number_Promotions (3,651): This variable has the highest Shapley number, suggesting that it is the most influential characteristic on attrition decisions. A higher number of promotions seems to be associated with a lower probability of attrition.
- NumeroRegulares_shap (2,867): The number of regular courses passed also has a significant influence on students' decision-making. A higher number of regular courses passed is associated with a lower probability of dropout.
- Average_grades_shap (2,569): Grade point average is another relevant variable in decision making. A higher GPA seems to be an important factor in student retention.

- TimeFaculty_shap (2.533): The length of time a student has been in faculty is also an influential variable. The longer a student has been at school, the less likely he/she is to drop out.
- AgeLastActivity_shap (2,510): The age at which a student undertook his or her last academic activity has a significant impact. Younger students in their last activity tend to have a lower probability of dropping out.

These five factors listed above (number of promotions, number of regular courses passed, grade point average, time in college and age at last activity) emerge as the most influential predictors of student attrition, according to the Shapley numbers. These findings can be valuable in identifying at-risk students and developing targeted interventions to improve retention. The graphs with the Shapley Values for the most important variables can be seen below.

Figure 1. Shapley Number values for TimeFaculty

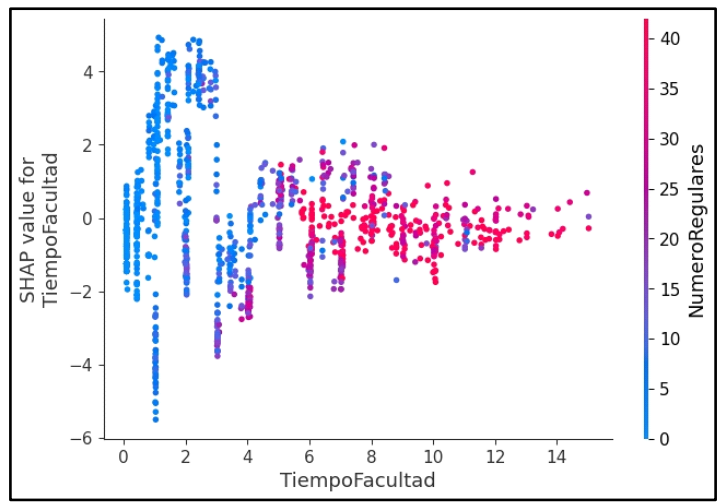

Figure 2. Shapley Number values for AgeLastActivity

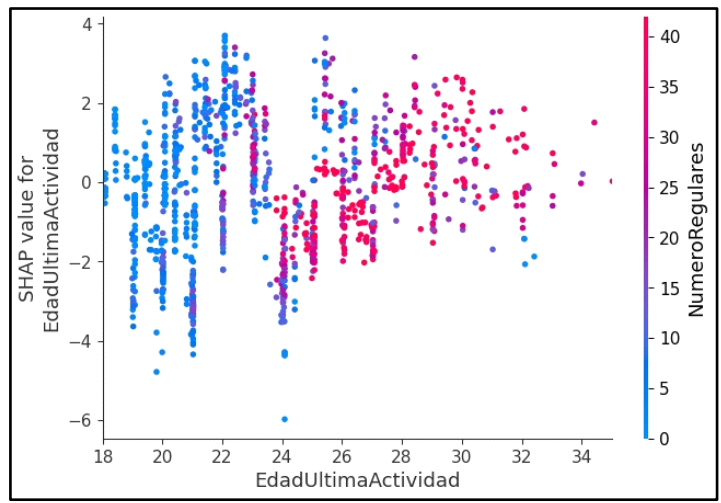

Figure 3. Shapley Number Values for RegularNumbers

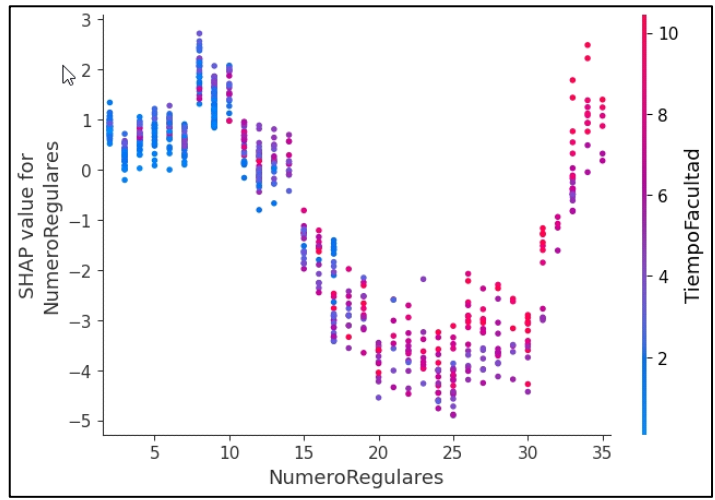

Figure 4. Shapley Number values for ApprovedNumbers

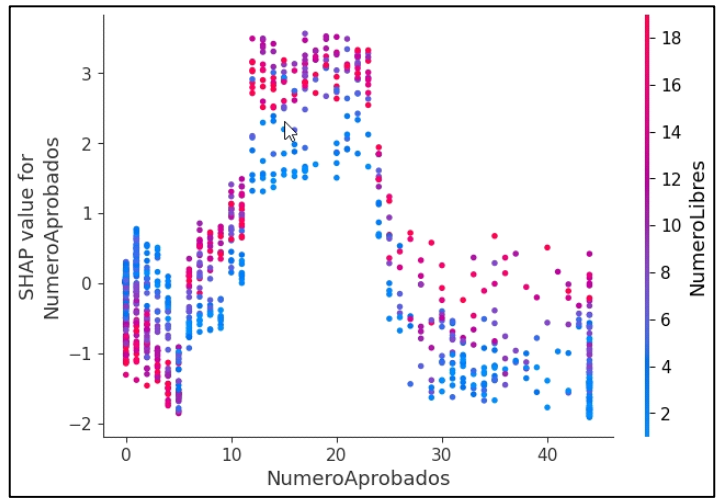

Figure 5. Shapley Number Values for NumberPromotions

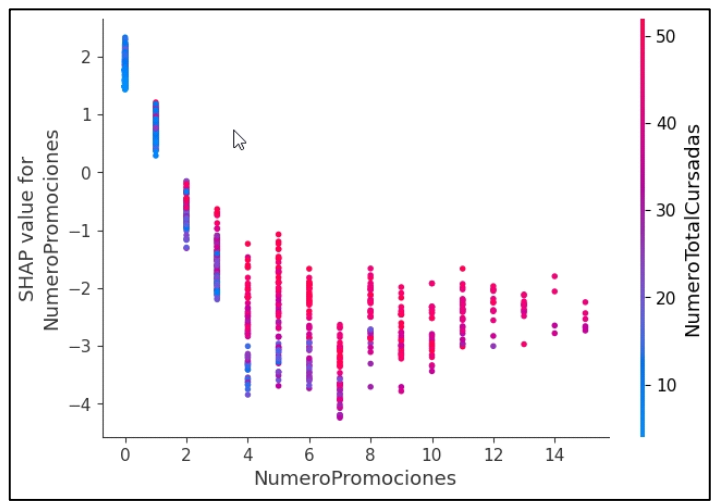

Figure 6. Shapley Number Values for NumberReproved

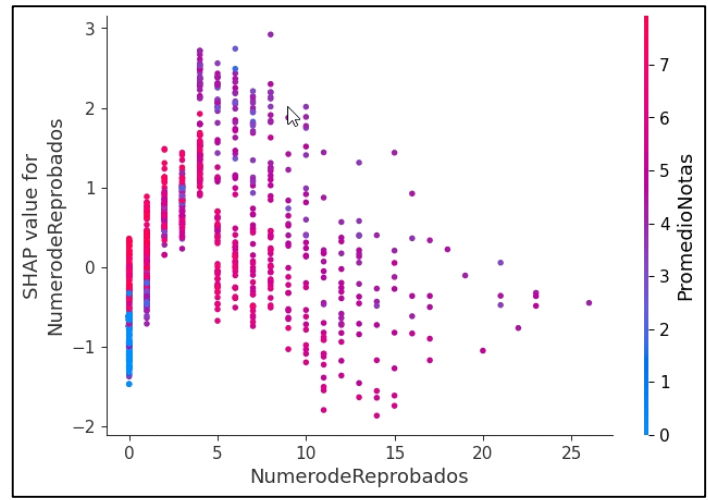

## 4. Discussion

### 4.1. Length of time spent in studies.

The data presented show the relationship between the time a student has stayed at the faculty **(TimeFaculty) and the influence of this variable on the probability of dropping out (ShapvalueDropout).** Since Shapley values greater than 0 indicate a higher probability of dropping out, while values less than 0 suggest a higher probability of staying, we can make the following observations:

- Initial Times (0 to 3 years): For students who have been in school for a relatively short period of time (0 to 3 years), the values of Shap_value_Dropout are mainly negative. This indicates that, in this initial period, a longer time at school is associated with a lower probability of dropping out. Negative values suggest that this variable contributes positively to the probability of staying in this time range.

- Intermediate Times (3 to 6 years): In the range of 3 to 6 years of faculty tenure, the Shap_value_Leave values fluctuate, but the general trend is towards negative values. This suggests that, in this intermediate period, time at the faculty remains a positive contributing factor to the probability of tenure, although the influence may not be as strong as in the early years.

- Advanced Times (6 years onwards): For students who have been in the faculty for more than 6 years, the values of Shap_value_Dropout tend to be positive. This indicates that, after a certain point of faculty tenure, a longer time is associated with a higher probability of dropping out. The positive influence on the probability of dropping out becomes more pronounced as students continue their time at the faculty.

The analysis of Shapley's values for the variable "FacultyTime" suggests a relevant pattern. Students who have been in school for a relatively short or moderate amount of time have a higher probability of st higher probability of dropping out. This suggests that there is a tipping point at which time at school ceases to have a positive effect on the probability of staying and starts to have a negative influence on the probability of dropping out.

**4.2. Student's age at last recorded activity**

The data presented show the relationship between the age at which a student last engaged in academic activity (Age_last_activity) and the influence of this variable on the probability of dropping out (Shap_value_leaving). Since Shapley values greater than 0 indicate a higher probability of dropping out, while values less than 0 suggest a higher probability of staying, we can make the following observations:

- Starting ages (18 to 19 years): For students who did their last academic activity at a young age (between 18 and 19 years), the values of Shap_value_Dropout are varied. Some ages within this range have positive values, indicating a higher probability of dropping out, while others have negative values, suggesting a higher probability of staying. This could be due to other variables not considered in this analysis that influence students' decision making.

- Intermediate Ages (20 to 26 years): In the age range 20 to 26 years, the Shap_value_Dropout values are predominantly positive. This indicates that, in general, as students get older in this range, they have a higher probability of dropping out. The positive influence on the probability of dropping out becomes more pronounced as age increases within this range.

- Advanced ages (27 years and older): For students who did their last academic activity at advanced ages (27 years and older), the values of Shap_value_Dropout are mainly positive and tend to increase. This suggests that, after a certain age, greater "ageing" is associated with a higher probability of dropping out.

The analysis of the Shapley values for the variable "Age_last_activity" indicates that the age at which a student undertook his or her last academic activity has an influence on the probability of dropping out. In general, younger ages (18 to 19 years) and older ages (27 years and older) are associated with a higher probability of dropping out, while intermediate ages (20 to 26 years) tend to show a positive influence on the probability of dropping out with increasing age. These findings suggest that age may be a relevant factor in students' decisions about staying in school.

**4.3. Courses taken and approved.**

The data presented show the relationship between the number of regular courses passed (NumeroRegulares) and the influence of this variable on the probability of dropping out (Shap_value_Dropout). Since Shapley values greater than 0 indicate a higher probability of dropping out, while values less than 0 suggest a higher probability of staying on, we can make the following observations:

- Low Number of Regular Courses (0 to 5): For students who have passed a low number of regular courses (0 to 5), the values of Shap_value_Dropout are positive, indicating a higher probability of dropping out. This suggests that students who have passed few regular courses have a higher propensity to drop out. As the number of regular courses passed increases within this range, the positive influence decreases, but remains significant.

- Moderate Number of Regular Courses (6 to 13): In the range of 6 to 13 regular courses passed, the Shap_value_Dropout values fluctuate, but tend to be positive overall. This suggests that even when students pass a moderate number of regular courses, there is still a positive influence on the probability of dropping out, although this influence becomes less pronounced as more courses are passed.

- High Number of Regular Courses (14 and above): For students who have passed a significantly high number of regular courses (14 onwards), the values of Shap_value_Dropout are negative. This indicates that a higher number of regular courses passed is associated with a lower probability of dropout. As students complete more regular courses, the negative influence on the probability of dropout increases significantly.

Analysis of the Shapley values for the variable "NumberRegulars" reveals interesting patterns. Students who have passed a low or moderate number of regular courses have a higher probability of dropping out, while those who have passed a significantly high number of regular courses have a lower probability of dropping out. These findings suggest that academic progress and completion of regular courses are crucial factors for student retention.

### 4.4. Number of subjects promoted

The data presented show the relationship between the length of time a student has stayed at the faculty (NumeroPromociones) and the influence of this variable on the probability of dropping out (Shap_value_Abandono). Since Shapley values greater than 0 indicate a higher probability of dropping out, while values less than 0 suggest a higher probability of staying, we can make the following observations:

- Initial Times (0 to 2 years): Students who have been in the faculty for a relatively short period of time (0 to 2 years) have values of Shap_value_Dropout close to 0. This suggests that, in this time interval, the variable "NumberPromotions" does not have a significant impact on the probability of leaving or staying. The values fluctuate around zero, indicating a neutral influence.

- Intermediate Times (2 to 5 years): As time in school increases in the range of 2 to 5 years, the values of Shap_value_Dropout begin to decrease, becoming negative. This suggests that, in this period, more time in school is associated with a lower probability of dropping out. Negative Shapley values indicate that this variable contributes positively to the probability of staying.

- Advanced Times (5 years onwards): For students who have been at the faculty for more than 5 years, the values of Shap_value_Dropout continue to be negative and decrease further. This reinforces the idea that time at school is an important factor in predicting tenure. The longer a student has been at school, the less likely he/she is to drop out.

The analysis of the Shapley values for the variable "NumberPromotions" reveals that this variable has a significant influence on the probability of dropping out. A longer time in the faculty is strongly associated with a higher probability of staying, while students who have been in the faculty for a shorter period do not show a clear influence of this variable on their dropout decisions. These findings are consistent with the intuition that faculty experience and continuity may have a positive impact on student retention.

**4.5. Number of subjects passed.**

The data presented show the relationship between the number of subjects passed (Number_cources_passed) and the influence of this variable on the probability of dropping out (Shap_value_Dropout). Since Shapley values greater than 0 indicate a higher probability of dropping out, while values less than 0 suggest a higher probability of staying on, we can make the following observations:

- Low Number of Subjects Passed (0 to 11): For students who have passed a low number of subjects (0 to 11), the values of Shap_value_Dropout are mainly negative. This indicates that, in general, a lower number of passed subjects is associated with a higher probability of dropping out. The negative influence on the probability of dropout is more pronounced in the range of 0 to 7 passed subjects.

- Moderate Number of Subjects Passed (12 to 27): In the range of 12 to 27 subjects passed, the values of Shap_value_Dropout fluctuate, but mostly tend to be positive. This suggests that as students pass a moderate number of subjects, the positive influence on the probability of dropping out becomes more prominent. Students who have passed between 12 and 27 subjects often have a higher probability of dropping out.

- High Number of Passed Subjects (more than 27): For students who have passed a significantly high number of subjects (more than 27), the values of Shap_value_Dropout are predominantly positive and tend to increase. This indicates that a higher number of passed subjects is strongly associated with a higher probability of dropping out. The positive influence on the probability of dropping out becomes more pronounced as the number of subjects passed increases.

The analysis of the Shapley values for the variable "Number_of_cources_passed" suggests that academic progress and the number of subjects passed have a significant influence on the probability of dropping out. In particular, students who have passed a low or moderate number of subjects tend to show a higher probability of dropping out, while those who have passed a significantly high number of subjects also have a higher probability of dropping out. These findings underline the importance of monitoring and supporting students' academic progress to improve retention.

**4.6. Number of Failed Subjects**

The data presented show the relationship between the number of subjects failed (Number_of_Courses_Failed) and the influence of this variable on the probability of dropping out (Shap_value_Dropout). Since Shapley values greater than 0 indicate a higher probability of dropping out, while values less than 0 suggest a higher probability of staying on, we can make the following observations:

- Low Number of Failed Subjects (0 to 5): For students who have a low number of failed subjects (0 to 5), the values of Shap_value_Dropout are mainly negative, indicating a lower probability of dropping out. This suggests that students who have failed few subjects tend to have a higher probability of staying in the faculty. Negative values are more pronounced in the range of 0 to 3 failed subjects.

- Moderate Number of Failed Subjects (6 to 16): In the range of 6 to 16 failed subjects, the Shap_value_Dropout values fluctuate, but tend to be positive overall. This indicates that, as the number of failed subjects increases within this range, the positive influence on the probability of dropping out becomes more prominent. Students who have failed between 6 and 16 subjects tend to show a higher probability of dropping out.

- High Number of Failed Subjects (more than 16): For students who have a significantly high number of failed subjects (more than 16), the values of Shap_value_Dropout are predominantly positive and tend to increase. This indicates that a higher number of failed subjects is strongly associated with a higher probability of dropping out. The positive influence on the probability of dropping out becomes more pronounced as the number of failed subjects increases.

The analysis of the Shapley values for the variable "Number_of_failed_Courses" suggests that the number of failed subjects has a significant influence on the probability of dropping out. Students who have failed few subjects tend to show a lower probability of dropping out, while those who have failed a moderate or high number of subjects tend to have a higher probability of dropping out. These findings emphasise the importance of providing additional support to students facing academic difficulties in order to improve retention.

**4.7. Limitations and bias considerations**

When considering the previously discussed results, it is essential to keep in mind some limitations and possible biases in our research. First, although we have identified significant variables that influence the likelihood of academic dropout, this study is based on historical data of civil engineering students from a specific institution, which may limit the generalisability of our findings to other academic contexts. In addition, we have not considered certain socio-economic or personal variables that might influence students' dropout decisions. Despite these limitations, our results provide valuable insight into key variables that should be considered by educational institutions when developing retention and student support strategies. Future research can address these limitations and further explore the complex interactions that influence academic dropout decisions.

It is essential to highlight that, although we carried out an exhaustive analysis using the SHAP (Shapley Additive Explanations) method, the results obtained do not establish a causal relationship between the variables analysed and academic dropout. In other words, we cannot infer that the variables identified and their respective outcomes are the direct causes of student dropout. What these results do provide are valuable and meaningful clues about the influences and patterns associated with the likelihood of dropout. Academic retention is a multidimensional and complex phenomenon, and students' decisions may be influenced by several interrelated factors, some of which may not have been considered in this analysis. Therefore, it is essential that future research further explores these relationships and considers possible additional variables to gain a more complete understanding of the determinants of academic dropout.

## 5. Conclusions

In summary, this study has shed light on the factors that influence the probability of academic dropout in civil engineering students at FACET-UNT. We have identified significant variables, such as the number of subjects failed, the number of subjects passed, the age at which the last academic activity took place and the length of time spent at the faculty, which are associated with the probability of dropping out. However, it is important to emphasise that these results do not establish causal relationships but offer valuable clues to better understand this complex phenomenon. In addition, we have used the SHAP method to analyse the relative importance of these variables in dropout decision-making, highlighting the need for more personalised support for students facing academic difficulties. While this study contributes to our understanding of student retention, it is critical to recognise the inherent limitations and lack of causality in our findings. Educational institutions can use this information as a starting point to develop more effective retention strategies and provide more targeted support to students.

The practical implications derived from the findings of this study are significant for educational institutions wishing to improve their student retention strategies. First, since variables such as the number of subjects failed, the number of subjects passed, the age at which the last academic activity took place and the length of time spent in the faculty have an impact on the likelihood of dropout, institutions can use this information to identify at-risk students and provide them with personalised support. This could include academic counselling, targeted tutoring for problem areas and follow-up programmes to monitor academic progress closely.

Furthermore, the application of the SHAP method to analyse the relative importance of these variables highlights the need to consider more advanced approaches to data analysis and modelling in retention decision-making. Institutions can take advantage of similar techniques to better understand the factors influencing dropout and design more effective interventions.

For future research, it is recommended to broaden the scope of this study and consider including additional variables, such as socio-economic factors. This will allow for a more complete understanding of the determinants of dropout. In addition, longitudinal research that follows students throughout their academic trajectory could provide valuable information on how these factors evolve over time.

Ultimately, this study underscores the importance of addressing student retention as a multidimensional challenge that requires holistic and personalised approaches. Evidence-based recommendations and the application of advanced data analysis techniques can contribute significantly to improving retention rates and ultimately to students' academic success.

**CONFLICTS OF INTEREST**

The author declares that he has no competing financial interests or known personal relationships that could have influenced the manuscript presented in this article.
**ETHICS COMMITTEE APPROVAL TO CONDUCT RESEARCH**

<div style="text-align:right">San Miguel de Tucumán, 12 November 2020.<br>Expiry Ref. NO. 60.183/2020</div>

 -HAVING SEEN: the presentation made by Prof. Hugo Roger Paz, Associate Professor of Basic Hydraulics, by which he requests authorization to develop his thesis in the Department of Construction and Civil Works of the FACET; and WHEREAS: Prof. Roger Paz, is pursuing a Doctorate in Education at the Faculty of Philosophy and Letters of the UNT; That to develop his thesis work he must conduct anonymous surveys to teachers and students of the career. Domingo Sfer, Head of the Department of Construction and Civil Works, endorses this request; Therefore, THIS DECANATE AUTHORISES Prof. Hugo Roger Paz, Associate Professor of Basic Hydraulics, to develop his thesis within the Department of Construction and Civil Works of the FACET.————————————————-